\documentclass{aa}  
\usepackage{graphicx}
\usepackage{txfonts}
\usepackage[colorlinks=true, linkcolor=blue, citecolor=blue, filecolor=blue, urlcolor=blue]{hyperref}

\begin{document}

   \title{Identifying multiplets of IceCube alert events}


   \author{M. Karl
          \inst{1, 2}
          \and
          P. Padovani\inst{2}
          \and
          P. Giommi\inst{3,4,5}
            }

    \institute{Technische Universit\"at M{\"u}nchen, TUM School of Natural Sciences, Physics Department, 
James-Frank-Str. 1, D-85748 Garching bei M{\"u}nchen, Germany\\
\email{martina.karl@tum.de}
            \and
            European Southern Observatory, Karl-Schwarzschild-Straße 2, 85748 Garching bei M\"unchen, Germany
            \and
            Institute for Advanced Study, Technische Universität M\"unchen,
Lichtenbergstrasse 2a, D-85748 Garching bei M\"unchen, Germany
            \and
            Center for Astrophysics and Space Science (CASS), New York University Abu Dhabi, PO Box 129188 Abu Dhabi, United Arab Emirates
            \and    
            Associated to INAF, Osservatorio Astronomico di Brera, via Brera, 28, I-20121 Milano, Italy 
    }

   \date{Received ....; accepted ....}

 
  \abstract
   {The IceCube Neutrino Observatory publishes ``alert events'', i.e. detections of high-energy neutrinos with a moderate-to-high probability of being of astrophysical origin. While some events are produced in the atmosphere, a fraction of alert events should point back to their astrophysical sources.}
   {We aim to identify multiple alert events possibly related to a single astrophysical counterpart by searching for spatial and temporal clusterings in 13 years of alert data.}
   {We identify spatial clusters (``multiplets'') by checking for events overlapping within their uncertainty regions. In order to reduce chance coincidences and to improve the signal purity of our sample, we apply different thresholds. We investigate the weighted mean position of these multiplets for an over-fluctuation of $\gamma$-ray counterparts. As a final step, we apply expectation maximization to search for temporal clusters around the identified weighted mean positions.}
   {We find no statistically significant clustering of alert events around a specific origin direction or in time.}
   {This could be because the selections are still dominated by atmospheric background. Another possibility is that we are not yet sensitive enough and only detect single events from sources. In this case, we need more data in order to observe a clustering of events around their origin.}


   \maketitle
%

\section{Introduction}
The IceCube Neutrino Observatory detects astrophysical neutrinos of mostly unknown origin \cite[for example,][]{2022ApJ...928...50A}. So far, only three sources have been identified: the blazar TXS~0506+056 \citep{icfermi}, the Seyfert Type II and starburst galaxy NGC~1068 \citep{doi:10.1126/science.abg3395f}, and the 
Galactic plane \citep{GalacticPlane2023}. There are hints for a more general population of sources \citep[for example,][]{abbasi2024icecubesearchneutrinoemission}, which have yet to be confirmed. The first ever non-stellar neutrino source TXS~0506+056 was identified with the help of a high-energy event \citep{icfermi}. While below 200~TeV events are dominated by the atmospheric background, the highest-energy events are expected to be dominated by astrophysical neutrinos \citep{2022ApJ...928...50A}. Hence, selecting these events should provide a relatively signal-pure sample. Whenever the IceCube Collaboration observes an astrophysical neutrino candidate with a good spatial resolution and high reconstructed energy, they issue notifications via the General Coordinates Network (GCN) as GCN Notices\footnote{\url{https://gcn.nasa.gov/notices}} and GCN Circulars\footnote{\url{https://gcn.nasa.gov/circulars}} to alert the astronomical community and trigger follow-up observations by other telescopes. Thus, these high-energy events are also referred to as ``alert events''.

A clustering of these signal pure events around a direction could point to astrophysical counterparts or at least indicate a common production site of clustered events. The lack of such clusters might suggest that either the sources are not as common as expected and we are dominated by atmospheric background, or that the sources are weak and emit only single events. It could also be that the mechanisms producing these neutrinos are different from what is currently understood. Previous work has derived constraints on the density and luminosity of steady standard candle neutrino sources dominating the high-energy ($\gtrsim 100$ TeV) neutrino flux
detected by IceCube \citep[for example,][]{Murase_2016}. 

However, the criteria to identify alert events have been revised and updated in 2019 \citep{blaufuss2019generation}, and an updated and revised collection of IceCube's highest-energy tracks observed between 2011 and 2023 is published in \citet{abbasi2023icecat1}. More recent alert events can be found in the GCN Notices and Circulars. Due to the longer time span and different selections, the number of released alerts has increased significantly. This new selection of alert events differs from the previously published alerts prior to 2019 \citep[see for example][]{Karl2023,Karl2024} and limits and constraints based on pre-revision alert selections might have to be revised as well. 

We investigate the non-detection of doublets and multiplets (for example by \citealt{Murase_2016}) with the revised and updated alert selection based on \citet{abbasi2023icecat1} and add more recent GCN Notices and Circulars. We look for overlapping events (``multiplets'') with the aim of identifying clustering alert events emitted by astrophysical neutrino sources. 

Since the production mechanisms of neutrinos also produce $\gamma$-ray emission, we take the mean multiplet positions and search for an excess in $\gamma$-ray-detected blazars. We do note that $\gamma$-ray emission can be attenuated and cascade down to lower energies. However, TXS~0506+056 as the template IceCube alert neutrino source does emit $\gamma$-rays. 
As a last step, we look for a temporal clustering of alert events. 

\section{IceCube alert events}\label{sec:alert_events}

The IceCube Neutrino Observatory constantly monitors the whole sky. This makes it ideally suited to alert other observatories of relevant detections and enable and trigger follow-up observations of potential transient phenomena. In 2016, the IceCube Collaboration started to publish high-energy events with a track-like signature in the detector nearly immediately after observation \citep{Aartsen_2017} in its realtime program. This realtime program was updated and revised in 2019 \citep{blaufuss2019generation}. Previous events, dating back to 2011, were revised in IceCat-1 \citep{abbasi2023icecat1}, which is a list of all highly energetic neutrinos with a track-like signature satisfying the updated alert criteria up to November 2023. There are two streams of alert events: the ``gold'' stream and the ``bronze'' stream. The gold stream provides an average astrophysical signal purity of $\approx50\%$ and the bronze stream an average signal purity of $\approx30\%$ \citep{blaufuss2019generation, abbasi2023icecat1}. 

The importance of high-energy neutrino events for the identification of astrophysical sources is emphasized by the first observation of a non-stellar astrophysical neutrino source. This source was indeed identified after the detection of an extremely highly energetic neutrino event (IC170922A, now belonging to the gold stream) that pointed back at the blazar TXS~0506+056 \citep{icfermi}, which was flaring in $\gamma$-rays at the time of the neutrino detection. IceCube observes $\sim 11$ neutrino events of gold classification per year \citep{abbasi2023icecat1}.

Once the realtime system identifies highly energetic neutrino events, a first notification in the General Coordinates Network (GCN) goes out as a GCN Notice 
to the astrophysical community \citep{Aartsen_2017, blaufuss2019generation}. This first GCN Notice includes, among other information, the origin direction, uncertainty area, and estimated neutrino energy based on a fast and simple reconstruction algorithm. A few hours later, after more sophisticated and time-intense algorithms are completed, an update to the reconstructed directions and uncertainties is issued as a GCN Circular.  
For calculating the reconstructed neutrino energy, the IceCube Collaboration assumes an underlying power-law emission of astrophysical sources of alert events, following $ \propto E^{-2.19} $ \citep{abbasi2023icecat1}. Currently, the IceCube Collaboration provides the reconstruction values as reported in the GCN Circulars as the final reconstruction, as these are the quantities reported in IceCat-1. However, \cite{Sommani:2023VW} concluded that there are different reconstruction algorithms, such as the one used for the GCN Notices, which provide reliable reconstructions with smaller uncertainty areas and are less affected by known systematic effects. 

In \citet{2024GCN.38267....1I}, the IceCube Collaboration introduces an update to the ``follow-up'' reconstruction issued after the first GCN Notices, starting with the IceCube alerts published end of September 2024. This update aims to improve the angular uncertainties and their coverage, which should help to identify the correct counterpart and spatial clusterings of alert events \citep{2024GCN.38267....1I}. Since this update is applied to events issued at the end of September 2024 and later, the majority of published alert events to this date (December 2024) do not have updated contours. 

Combining the gold and bronze stream, the IceCube Collaboration has published 348 alert events by November 2023\footnote{\url{https://dataverse.harvard.edu/dataset.xhtml?persistentId=doi:10.7910/DVN/SCRUCD}}. After removing alerts flagged as probable cosmic ray events and adding alert events published until beginning of July 2024 based on GCN circulars, we get a final sample of 355 events. 

We look for overlapping events to identify clustering alert events emitted by astrophysical neutrino sources. Considering only the alert events without including a further (lower-energy) neutrino component is also motivated by \citet{Abbasi_2024}, which found no general connection of alert events to lower energetic neutrino emission. The case of TXS~0506+056, where there was an alert event and a neutrino flare at lower energies some years before the alert event \citep{neutrino}, is so far a unique case, and \citet{Abbasi_2024} found no similar cases. Hence, we expect the alert events to be the dominant signature of their sources. This also agrees with the further identified non-stellar neutrino sources (apart from TXS~0506+056) to date: the Seyfert Type-II and starburst galaxy NGC~1068 \citep{doi:10.1126/science.abg3395f}, and the Milky Way (as a diffuse source) \citep{GalacticPlane2023}. In both cases, the signal was found by analyzing neutrino data going down to lower energies, and alert events did not contribute to the neutrino signal. Hence, we expect different contributions, emission processes, 
and source populations for the astrophysical diffuse neutrino fluxes at lower and higher energies, as, for example, proposed in \citet{Padovani_2024}.

\section{Reported multiplets}

\cite{IceCube_multiplet_2017} reported on a rare IceCube neutrino multiplet that was part of a different (optical) alert stream. These optical alerts are not available publicly but are issued directly to observatories when IceCube detects events within 100~s and within $3.5^\circ$ of each other \citep[see for example][]{2012A&A...539A..60A, 2015ApJ...811...52A, 2016_followups, Aartsen_2017}. Hence, these multiplet events differ from alert events based on criteria in \citet{blaufuss2019generation, abbasi2023icecat1}. Since these optical follow-up events are not publicly available, they are not included in this work. 

\citet{2024A&A...687A..59G} approached the multiplet questions slightly differently, by looking for $\gamma$-ray sources spatially coincident with IceCube alert events. They investigated different selections of alert events, combining, in all cases, alert events pre-revision with alert events post-revision. As mentioned in \citep{blaufuss2019generation, abbasi2023icecat1}, the criteria for pre-revision alert events differ from the criteria used for selecting post-revision alert events. Depending on their selection, \citet{2024A&A...687A..59G} identified either 14 $\gamma$-ray sources spatially coincident with two IceCube alert events, or 13 $\gamma$-ray sources spatially coincident with two IceCube alert events with ten additional $\gamma$-ray sources coincident with three IceCube alert events, and one $\gamma$-ray source coincident with four IceCube alert events. In all cases, the coincidences were consistent with chance associations and were not significant \citep{2024A&A...687A..59G}. 

\citet{sommani2024100tevneutrinoscoincident} reported two 100~TeV neutrino alert events (IC220424A \& 
IC230416A) from the direction of NGC~7469 with a chance coincidence of $3.3\,\sigma$. However, they use the reported values of the GCN Notices\footnote{\url{https://gcn.gsfc.nasa.gov/notices_amon_g_b/136565_2186969.amon} \\ \url{https://gcn.gsfc.nasa.gov/notices_amon_g_b/137840_57034692.amon}} with the first preliminary reconstruction of the neutrino event. When they calculate the chance probability using the values reported by the more sophisticated and more time-intensive algorithm in the GCN Circulars, the significance disappears \citep{sommani2024100tevneutrinoscoincident}. 

In our case, we want to conduct a statistical search for an accumulation of only revised alert events, such that the selection criteria remain consistent. We also investigate the whole alert sample, not just alert events at the precise location of a specific source as in \citet{sommani2024100tevneutrinoscoincident}. For this, we rely on a large enough number of alert events. Taking only the published GCN circular values reduces our number of alerts significantly. We start with the official IceCube alert catalog IceCat-1 and add alert events circulated since the latest update of IceCat-1 and until the beginning of July 2024. As an additional test, we adopt the strategy of \citet{sommani2024100tevneutrinoscoincident} and test alert events with the reconstruction values issued in the first GCN Notices.

\section{Multiplets for different area, energy, and signalness thresholds}

\begin{figure}
    \centering
    \includegraphics[width=\hsize]{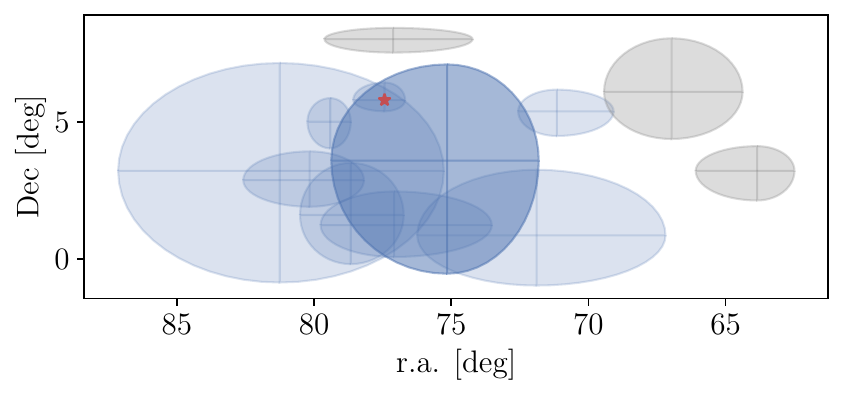}
    \caption{The neutrino alert sky in right ascension (r.a.) and declination (Dec) in the vicinity of TXS~0506+056 (red star). The regions show alert events with their respective uncertainty ellipses (90\% confidence level uncertainties). The central dark blue alert is identified as the ``center'' of the multiplet with the most overlapping events in this cluster. The grey alerts are not part of the multiplet.}
    \label{fig:TXS_alerts_multiplets}
\end{figure}

We identify a multiplet if the uncertainty regions of two or more alert events touch or overlap. We then count how often we find overlapping events. In cases where one alert spatially overlaps with several other alert events, we consider the alert event with the most multiplets and reject the remaining events. As an example, we consider the alerts displayed in Figure \ref{fig:TXS_alerts_multiplets}. When counting the number of overlapping events for each alert, we count, for example, seven overlapping events for the leftmost alert event. The central alert marked in dark blue, however, has eight overlapping events (including the leftmost alert event). We require one alert to contribute only to one multiplet, so we reject the multiplet of seven and keep the multiplet of eight centered around the dark blue alert. The alerts marked in grey do not contribute to the multiplet because they do not overlap with the central dark blue alert. 

We repeat this procedure for all alerts and count how often we find multiplets of 1, 2, 3, ... events. To determine if there is an over-fluctuation of multiplets, we generate a random neutrino alert sky by assigning random right ascension values\footnote{Due to IceCube's unique location directly at the South Pole, background events are uniformly distributed over right ascension when integrating over time periods greater than one day. IceCube's effective areas, however, depend on the zenith angle of events and we consider these dependencies by preserving the declination values of events.} to the alert events. Then, we count how many multiplets we see for the randomized alert events. To get a distribution of the expected number of background multiplets, we repeat this procedure $10^3$ times. As a next step, we compare the background expectation with the actual number of alert multiplets and assess the significance. If the actual number of multiplets exceeds a significance of 1\%, we increase the number of background realizations to guarantee a proper evaluation. We attempt to improve the signal purity of the alerts by testing different thresholds of maximally allowed uncertainty areas, minimally required alert energies, and minimally allowed signalness (a quantity published with each alert event assessing an event's probability to be astrophysical, see \citealt{blaufuss2019generation, abbasi2023icecat1}) and repeat the multiplet count for each selection. 

We select the area thresholds in descending order as the area corresponding to an equivalent radius of $3~\mathrm{deg}~(= 28.27~\mathrm{deg}^2$, following \citealt{Giommidissecting}), the mean alert uncertainty area ($21.77~\mathrm{deg}^2$), the 68\% quantile of all alert uncertainty areas ($12.63~\mathrm{deg}^2$), and the median alert uncertainty area ($6.27~\mathrm{deg}^2$). For each area threshold, $A_\mathrm{thresh}$, we select alert events with uncertainty areas $\leq A_\mathrm{thresh}$ and search for multiplets and their significance as described above.

Since the reconstructed neutrino energy can indicate how likely an event is astrophysical, we repeat the multiplet search for increasing energy thresholds, $E_\mathrm{thresh}$. However, as mentioned in \citet{abbasi2023icecat1}, the reconstructed neutrino energy is calculated assuming an underlying power-law emission following $\propto E^{-2.19}$. Changing the source emission spectrum could hence affect the reconstructed neutrino energies. For this work, we adopt the published energies based on the power-law assumption. We start with no energy threshold (with the lowest reconstructed alert energy of 54~TeV), including the complete alert sample, and then we apply the mean, the median, and the 68\% quantiles as thresholds. Similar to the previous procedure, we select alert events with energies above or equal to the respective energy threshold and search for multiplets as described above. 

However, by simply using the reconstructed neutrino energy, we do not consider detection efficiencies based on the detector effective area for different energies and declinations. With the GCN Notices IceCube also publishes the ``signalness'', which quantifies the probability of each event to be of astrophysical origin \citep{abbasi2023icecat1}. This quantity includes detector dependencies on declination and energy, but it also assumes an astrophysical energy spectrum $\propto E^{-2.19}$ \citep{abbasi2023icecat1}. Hence, the signalness will change when assuming a different energy spectrum. For this test now, we use the published signalness values and apply thresholds of the median, the mean, and the 68\% quantile.

We list all alert multiplets for different area, energy, and signalness thresholds in Tables \ref{tab:multiplets_area}, \ref{tab:multiplets_energy}, and \ref{tab:multiplets_signalness}. We find no significant over-fluctuation of multiplets for any energy, signalness, or area threshold.

\begin{figure}
    \centering
    \includegraphics[width=\hsize]{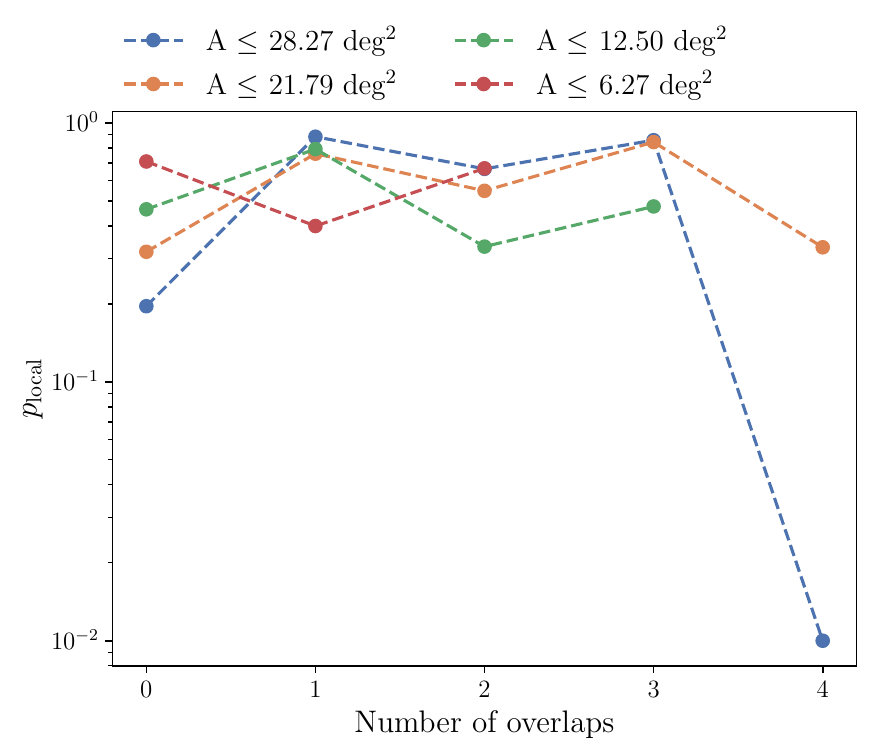}
    \caption{Local p-values for different area thresholds. Zero overlaps correspond to ``single'' alert events, one corresponds to a doublet (one alert event overlapping with another), and so forth.}
    \label{fig:pval_area}
\end{figure}

\begin{figure}
    \centering
    \includegraphics[width=\hsize]{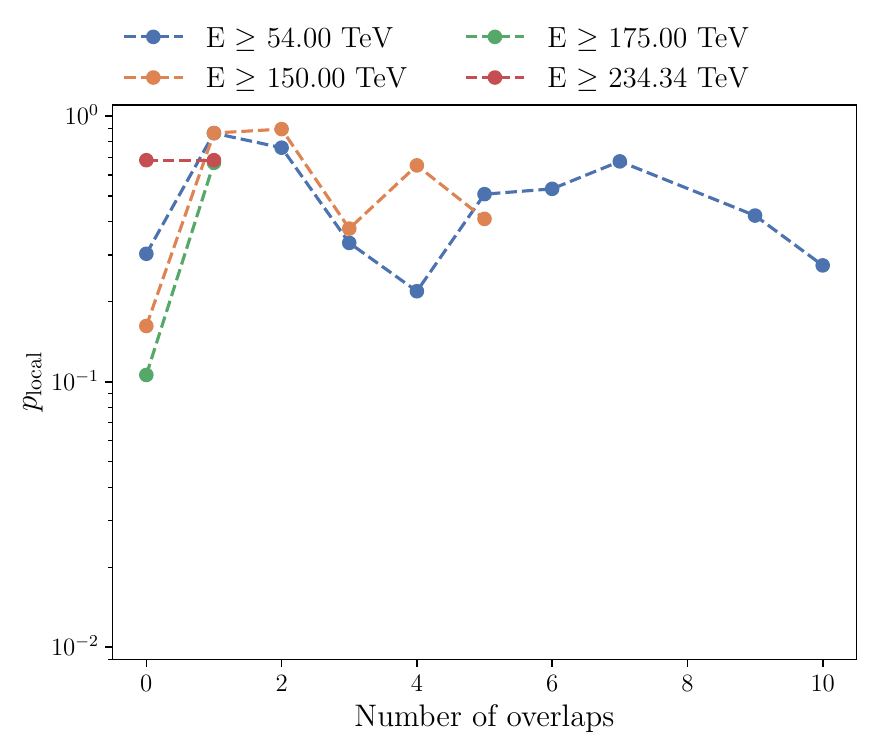}
    \caption{Local p-values for different energy thresholds. Zero overlaps correspond to ``single'' alert events, one corresponds to a doublet (one alert event overlapping with another), and so forth. The lowest energy threshold of 54~TeV includes the full alert sample since 54~TeV is the lowest alert energy in our selection.}
    \label{fig:pval_energy}
\end{figure}

\begin{figure}
    \centering
    \includegraphics[width=\hsize]{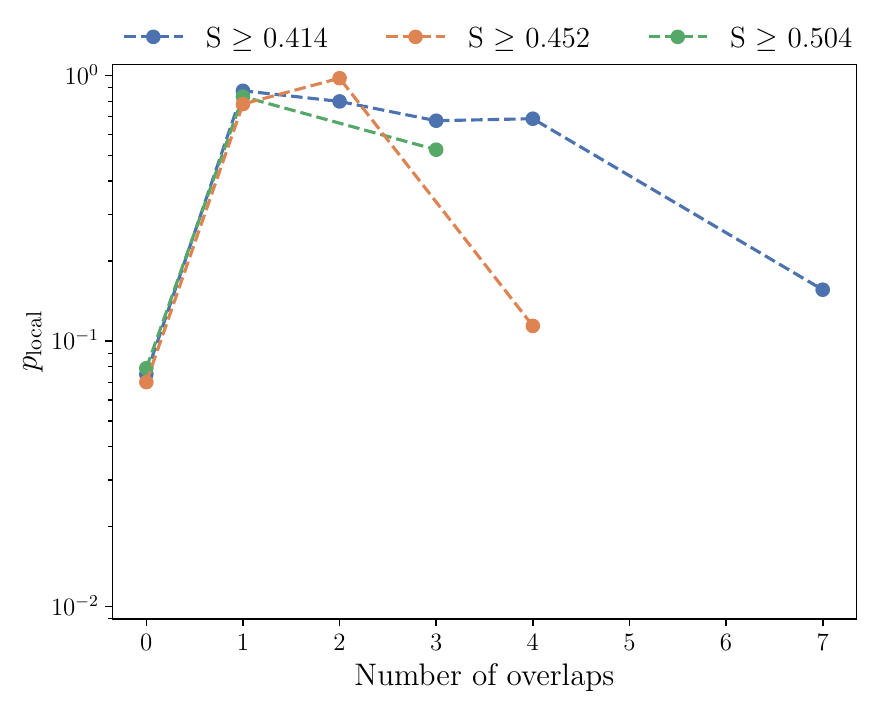}
    \caption{Local p-values for different signalness thresholds. Zero overlaps correspond to ``single'' alert events, one corresponds to a doublet (one alert event overlapping with another), and so forth. }
    \label{fig:pval_signalness}
\end{figure}

The most significant local p-value for a multiplet corresponds to $1\%$ for the area search ($A_\mathrm{thresh} = 28.27~\mathrm{deg}^2$), 11\% for the energy search ($E_\mathrm{thresh} = 175$~TeV), and 11\% for a signalness threshold of 0.452. We show the respective local p-values per number of multiplets in Figures \ref{fig:pval_area}, \ref{fig:pval_energy}, and \ref{fig:pval_signalness} for all area, energy, and signalness thresholds. These significances are not yet corrected for scanning multiple thresholds, which would decrease them
even further.

\longtab{
\begin{longtable}{lcccccl}
\caption{List of alert multiplets for different areas. We list the multiplets for the respective area threshold, starting with the smallest area. Multiplets with smaller threshold areas are also multiplets for larger area thresholds but are not repeated in the table. In the cases where increasing the area adds events to previously identified multiplets, we mark all multiplet events in bold font. The alert events' names are as reported in IceCat-1, with the respective right ascension (r.a.), declination, reconstructed neutrino energy, uncertainty area, signalness, and the alert classification. Alert events where the classification is only ``gold'' or ``bronze'' are not part of IceCat-1 (as of July 2024).}\label{tab:multiplets_area}\\
\hline\hline
 Name & r.a. & Dec & Energy & Area  & Signalness & Classification \\
 & [deg] & [deg] & [TeV] & [deg$^2$] & & \\
\hline
\endfirsthead
\caption{continued.}\\
\hline\hline
 Name & r.a. & Dec & Energy & Area  & Signalness & Classification \\
 & [deg] & [deg] & [TeV] & [deg$^2$] & & \\
\hline
\endhead
\hline
\endfoot
\hline
\hline
\noalign{\smallskip}
\multicolumn{7}{c}{6.27 deg$^2$} \\
\noalign{\smallskip}
\hline
IC110807A & $336.8^{+1.36}_{-1.98}$ & $1.53^{+0.93}_{-0.78}$ & 108.0 & 4.48 & 0.266 & gfu-bronze \\
IC140114A & $337.59^{+0.57}_{-0.92}$ & $0.71^{+0.97}_{-0.86}$ & 54.0 & 2.14 & 0.335 & hese-bronze \\
IC221224A & $335.74^{+0.97}_{-0.7}$ & $1.42^{+0.3}_{-0.37}$ & 111.0 & 0.88 & 0.278 & gfu-bronze \\
\hline
IC110902A & $9.76^{+2.86}_{-1.32}$ & $7.59^{+0.87}_{-0.86}$ & 243.0 & 5.63 & 0.609 & gfu-gold \\
IC140603A & $9.71^{+0.62}_{-0.88}$ & $7.56^{+0.53}_{-0.83}$ & 152.0 & 1.59 & 0.377 & gfu-bronze \\
\hline
IC110930A & $267.01^{+1.19}_{-1.14}$ & $-4.44^{+0.6}_{-0.79}$ & 160.0 & 2.54 & 0.429 & gfu-bronze \\
IC220205B & $267.01^{+0.67}_{-0.56}$ & $-3.66^{+0.41}_{-0.5}$ & 216.0 & 0.88 & 0.595 & gfu-gold \\
\hline
\end{longtable}
}

\longtab{
\begin{longtable}{lcccccl}
\caption{List of alert multiplets for different energy thresholds. We list the multiplets for the respective energy threshold, starting with the largest energy. The lowest energy includes the full alert sample and corresponds to no cuts at all. Multiplets with larger threshold energies are also multiplets for smaller energy thresholds but are not repeated in the table. In the cases where lowering the energy threshold adds events to previously identified multiplets, we mark all multiplet events in bold font. The alert events' names are as reported in IceCat-1, with the respective right ascension (r.a.), declination, reconstructed neutrino energy, uncertainty area, signalness, and the alert classification. Alert events where the classification is only ``gold'' or ``bronze'' are not part of IceCat-1 (as of July 2024).}\label{tab:multiplets_energy}\\
\hline\hline
 Name & r.a. & Dec & Energy & Area & Signalness & Classification \\
 & [deg] & [deg] & [TeV] & [deg$^2$] & & \\
\hline
\endfirsthead
\caption{continued.}\\
\hline\hline
 Name & r.a. & Dec & Energy & Area  & Signalness & Classification \\
 & [deg] & [deg] & [TeV] & [deg$^2$] & & \\
\hline
\endhead
\hline
\endfoot
\hline
\hline
\noalign{\smallskip}
\multicolumn{7}{c}{234.34 TeV} \\
\noalign{\smallskip}
\hline
IC181023A & $270.18^{+1.89}_{-1.71}$ & $-8.42^{+1.13}_{-1.55}$ & 237.0 & 7.50 & 0.153 & ehe-gold \\
IC220425A & $268.24^{+1.93}_{-1.58}$ & $-10.73^{+1.45}_{-1.67}$ & 604.0 & 8.45 & 0.169 & ehe-gold \\
\hline
\noalign{\smallskip}
\multicolumn{7}{c}{175.00 TeV} \\
\noalign{\smallskip}
\hline
IC140927A & $50.89^{+3.91}_{-5.14}$ & $-0.63^{+1.49}_{-1.42}$ & 182.0 & 20.68 & 0.481 & gfu-gold \\
IC220304A & $48.78^{+8.09}_{-6.5}$ & $4.48^{+5.0}_{-6.25}$ & 263.0 & 128.52 & 0.631 & gfu-gold \\
\hline
IC170105A & $309.95^{+5.01}_{-7.56}$ & $8.16^{+2.0}_{-3.34}$ & 198.0 & 52.19 & 0.535 & gfu-gold \\
IC220306A & $314.82^{+0.53}_{-0.44}$ & $8.61^{+0.53}_{-0.45}$ & 413.0 & 0.74 & 0.774 & gfu-gold \\
\hline

\end{longtable}
}

\longtab{
\begin{longtable}{lcccccl}
\caption{List of alert multiplets for different signalness thresholds. We list the multiplets for the respective signalness threshold, starting with the largest signalness. Multiplets with larger threshold signalness are also multiplets for smaller signalness thresholds but are not repeated in the table. In the cases where lowering the signalness threshold adds events to previously identified multiplets, we mark all multiplet events in bold font. The alert events' names are as reported in IceCat-1, with the respective right ascension (r.a.), declination, reconstructed neutrino energy, uncertainty area, signalness, and the alert classification. Alert events where the classification is only ``gold'' or ``bronze'' are not part of IceCat-1 (as of July 2024).}\label{tab:multiplets_signalness}\\
\hline\hline
 Name & r.a. & Dec & Energy & Area  & Signalness & Classification \\
 & [deg] & [deg] & [TeV] & [deg$^2$] & & \\
\hline
\endfirsthead
\caption{continued.}\\
\hline\hline
 Name & r.a. & Dec & Energy & Area  & Signalness & Classification \\
 & [deg] & [deg] & [TeV] & [deg$^2$] & & \\
\hline
\endhead
\hline
\endfoot
\hline
\hline
\noalign{\smallskip}
\multicolumn{7}{c}{0.504} \\
\noalign{\smallskip}
\hline
IC130125A & $7.67^{+6.46}_{-5.92}$ & $74.14^{+3.36}_{-2.82}$ & 165.0 & 16.42 & 0.531 & gfu-bronze \\
IC140410A & $2.11^{+151.51}_{-58.92}$ & $81.22^{+8.0}_{-6.94}$ & 246.0 & 376.89 & 0.627 & gfu-gold \\
\hline
IC131124A & $285.16^{+2.2}_{-1.54}$ & $19.47^{+1.43}_{-1.46}$ & 180.0 & 8.00 & 0.553 & gfu-gold \\
IC220221A & $287.84^{+4.57}_{-4.39}$ & $20.74^{+2.22}_{-4.46}$ & 157.0 & 43.96 & 0.673 & hese-gold \\
\hline
IC150714A & $326.29^{+1.49}_{-1.32}$ & $26.36^{+1.89}_{-2.19}$ & 439.0 & 8.07 & 0.841 & gfu-gold \\
IC221210A & $332.58^{+10.46}_{-10.81}$ & $22.75^{+4.46}_{-8.55}$ & 165.0 & 200.43 & 0.508 & gfu-bronze \\
\hline
IC151013A & $178.72^{+1.11}_{-1.15}$ & $52.37^{+1.11}_{-1.11}$ & 156.0 & 2.41 & 0.516 & gfu-bronze \\
\hline
\end{longtable}
}

\subsection{$\gamma$-ray sources at mean multiplet positions}

\begin{figure}
    \centering
    \includegraphics[width=\hsize]{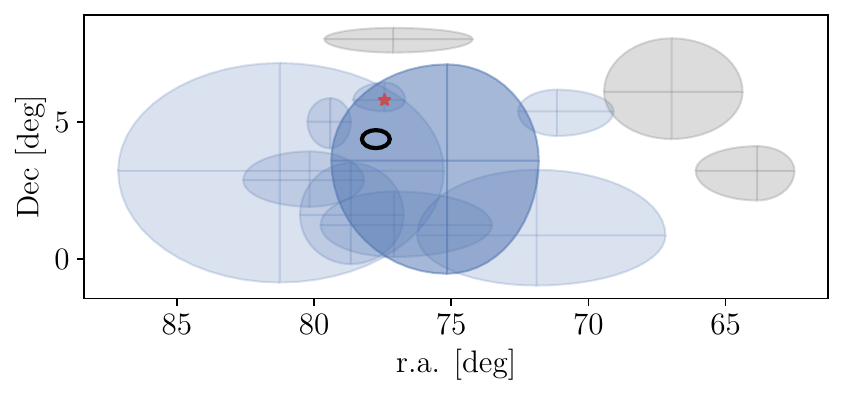}
    \caption{The same multiplet as in Fig. \ref{fig:TXS_alerts_multiplets}. The black ellipse shows the location and uncertainties on the weighted arithmetic circular mean position.}
    \label{fig:txs_multiplet_weighted}
\end{figure}

We expect the production of neutrinos to be accompanied by $\gamma$-rays. We investigate if the centers of our multiplets show a higher number of $\gamma$-ray-detected blazars compared to the average blazar density. For this, we calculate the weighted arithmetic circular mean positions and uncertainties for each multiplet. We use $1/\sigma_i^2$ as weight, with $\sigma_i$ as a vector of the mean right ascension and declination uncertainties. This reduces the investigated area drastically since the uncertainties on the weighted arithmetic circular mean positions are calculated with $\sigma_{\bar{x}} = \sqrt{1 / \sum_{i=1}^{n} \sigma_i^{-2}}$ (see for example the black ellipse in Figure \ref{fig:txs_multiplet_weighted}).

We then take the revised 4FGL blazar catalog (Giommi et al. 2024, in prep.) and count how many reported counterparts fall within the weighted mean area and compare this number with a background expectation. The revised 4FGL catalog excludes sources in the vicinity of the Galactic plane; hence, we also only include weighted positions in this search with $|b| > 10^\circ$. The background expectation results from the average number of objects per area in the revised 4FGL catalog. We find no excess of $\gamma$-ray-detected blazars in the weighted mean areas.

\subsection{Multiplets based on GCN Notices}
Following \citet{sommani2024100tevneutrinoscoincident}, we investigate how significant multiplets become when only considering the reconstruction properties issued in the first GCN Notices (with the 50\% error radius). This includes now only events that were issued in the realtime alert stream starting in mid-2019\footnote{\url{https://gcn.gsfc.nasa.gov/amon_icecube_gold_bronze_events.html}} (129 in total, after removing retracted alert events). We find one doublet (the one reported by \citet{sommani2024100tevneutrinoscoincident}, see Table \ref{tab:GCN_notice_doublets}) compatible with a chance probability of 47\%\footnote{\citet{sommani2024100tevneutrinoscoincident} got a $3.3\,\sigma$ significance by evaluating the spatial coincidence between the two neutrino alerts, IC220424A and IC230416A, and 
NGC 7469, whereas we investigate the significance of the multiplet for the whole sky.}. 

\begin{table*}
\caption{The only multiplet when taking the GCN Notices reconstruction values. The doublet was also reported by \citet{sommani2024100tevneutrinoscoincident}.}\label{tab:GCN_notice_doublets}  
\centering        
\begin{tabular}{lccccl}  
\hline\hline  
 Name & r.a. & Dec & 50\% Error & Energy & Classification \\
 & [deg] & [deg] & [deg$^2$] & [TeV]  & \\
\hline  
IC230416A & 345.83 & 9.01 & 0.20 & 127.29 & bronze \\
IC220424A & 345.76 & 8.86 & 0.26 & 183.99 & gold \\
\hline  
\end{tabular}
\end{table*}

\subsection{Discussion}
We do not find a significant spatial clustering of alert events around a common origin. 
This allows several interpretations. It might indicate that alert events are rare events and we need longer integration times to accumulate sufficient alerts from a source for a significant detection. It could also be that our selections and attempts to reduce chance associations were not sufficient, and a potential signal is hidden beneath the atmospheric background. When applying thresholds on uncertainty area, energy, and signalness, there are many aspects to consider. For example, the energy reconstructions require many intermediate steps and assume an underlying power-law emission with a defined spectrum $\left( \propto E^{-2.19} \right)$ of neutrino sources \citep{abbasi2023icecat1}. Such a power-law spectrum (with a negative spectral index) implies that high-energy events are accompanied by a larger flux of lower-energy events. However, \citet{Abbasi_2024} does not find such a correlation, and our approach of only considering alert events as a signal is not compatible with a power-law spectrum. A different energy spectrum, for example a harder neutrino spectrum, as suggested by \citet{Padovani_2024} and supported by the modeling of candidate neutrino sources in \citet{Rodrigues_etal_2024}, will most likely affect the reconstructed neutrino energy \citep{Rodrigues_etal_2024} and consequently change our selection when applying energy thresholds. 

Concerning the uncertainty areas, \citet{Sommani:2023VW} concluded that different reconstruction algorithms could reduce the uncertainty regions while providing reconstructed positions close to the true origin. \citet{gualda2021studiessystematicuncertaintyeffects} found that the published uncertainty regions do not always provide the expected coverage and might be larger or (for horizontal shallow events) smaller depending on the event properties. As mentioned in Section \ref{sec:alert_events}, the IceCube Collaboration announced an update to the muon track ``follow-up'' reconstruction issued after the first GCN Notice, starting with the IceCube alert IC-240929A\footnote{\url{https://gcn.gsfc.nasa.gov/notices_amon_g_b/139912_46959751.amon}, and \url{https://gcn.nasa.gov/circulars/37625}} \citep{2024GCN.38267....1I}. The update aims to improve the angular uncertainties and their coverage, which should help to identify the correct counterpart and spatial clusterings of alert events \citep{2024GCN.38267....1I}. Unfortunately, this update is applied to events issued at the end of September 2024 and later, whereas earlier events used for this study have not been updated. Different uncertainty regions would also affect our selection and potentially the number of multiplets we find for different area thresholds. This update affects only the second reconstruction published with the revised GCN Notices and Circular. When following the selection in \citet{sommani2024100tevneutrinoscoincident} by only considering the reconstructed values of the first GCN Notices (that remain unchanged with \citealt{2024GCN.38267....1I}), we also do not find a significant number of multiplets. This latter approach reduces the statistics of alert events to realtime events issued after mid-2019 and excludes a large part of events published in IceCat-1 where no GCN Notices are available. Hence, this attempt is probably limited by statistics. 

Considering our search for $\gamma$-ray detected blazars at the weighted circular mean position of identified multiplets, we do not find an excess of sources. In previous studies where a connection of $\gamma$-ray detected blazars and neutrinos were investigated, the authors usually searched within the alert uncertainty regions (or a scaled-up uncertainty area), in contrast to our approach in this work. For comparison, \citet{Giommidissecting} investigated mainly alert events prior to the revision described in Section \ref{sec:alert_events} and found a 3.2\,$\sigma$ correlation between intermediate-to-high-peaked blazars and neutrino alerts when increasing the uncertainty areas by a factor of 1.3. \citet{2024A&A...687A..59G} combined alert events before and after revision and did not find a significant spatial correlation of alert events and $\gamma$-ray counterparts. \cite{KouchEtAl2024} presented another association between blazars and IceCube alert events on a 2.17\,$\sigma$ level when enlarging the uncertainty areas by 1 degree in quadrature and taking blazars detected in the radio and optical of the CGRaBS catalog. These results emphasize furthermore that the association between counterparts and IceCube alert events relies heavily on the reconstructed uncertainties of the IceCube alert events, on top of the fact that the intrinsic association strength might be relatively small. 

We also note that \citet{Plavin_2020,10.1093/mnras/stad1467} found 2.9 and 3.4\,$\sigma$ associations between bright-radio blazars and IceCube events by adding $\sim 0.5^{\circ}$ to the published IceCube spatial uncertainties. The IceCube collaboration \citep{Abbasi_2023_correlation_search} could confirm the earlier result within a factor of 2 in the p-value (see their Table 4) although this was not the case when they used a more sophisticated description of the spatial
probability density function for the neutrino events and an updated event catalog.

\section{Time series analysis}

We now expand our search to possible temporal clusterings of alert events. For this, we use the unsupervised machine learning algorithm Expectation Maximization (EM) as presented in \citet{2024JCAP...07..057K, Abbasi_2024}. EM is based on a Gaussian mixture model, where we describe the signal, a temporal clustering, as Gaussian flares. We assume that there is a uniform background component with random events distributed over time, and alert events that are emitted during one or more Gaussian-shaped time windows as a signal. In the end, we compare two hypotheses: 

\begin{itemize}
    \item \textbf{Background Hypothesis}: There is no clustering in arrival time. We see $N$ uniformly distributed detection times. \\
    \item \textbf{Signal Hypothesis}: In addition to the uniform background, we observe temporally clustered alert events. We have $k$ neutrino flares with a certain strength expressed as the number of detected signal neutrinos, $n_{\mathrm{S,}k}$. The background component is then the remaining events not accounted for by the neutrino flares (so $N - \sum_k n_{\mathrm{S,}k}$ uniformly distributed detection times).
\end{itemize}

The likelihood comprising the signal and background probability is then maximized by EM (as described in \citealt{2024JCAP...07..057K, Abbasi_2024}). The result is a distribution of Gaussians with the best-fit values of their means $\mu_{\mathrm{T,}k}$, their widths, $\sigma_{\mathrm{T,}k}$, and their respective strengths, $n_{\mathrm{S},k}$. 

For each multiplet, we assume the weighted arithmetic circular mean position calculated in the previous section to be the position of a point-like neutrino source and then evaluate if we observe a temporal clustering of alerts close to that position. Hence, alert events close to that position should contribute more to a signal, whereas alert events more distant to that position are less likely to originate from that point-source position and should contribute less or not at all. To consider this when fitting neutrino flares, we assign weights based on the spatial and energy probability of each alert event to originate from our assumed source or from background. This weight is then applied to the time series as described in \citet{2024JCAP...07..057K, Abbasi_2024}. We calculate the weights for every alert on the sky and do not limit the contributing events to the multiplets because the multiplets consist only of a few events.

We describe the weight's spatial contribution by a Rayleigh distribution centered at the point-source position with the mean alert uncertainty as its spread (as defined in equation 3 of \citealt{Abbasi_2024}). Following \citet{2024JCAP...07..057K, Abbasi_2024}, we divide this signal probability by the background probability, where we describe the background by a uniform distribution over right ascension and a declination-dependent distribution based on the effective areas published in \citet{abbasi2023icecat1}. This ensures stronger weights for events close to the point-source position and very weak weights for events distant from the point-source position. To weigh events with higher energies accordingly, while taking detector effects (depending on energy $E$ and declination $\delta$) into account, we derive energy weights from the alerts' signalness. \citet{abbasi2023icecat1} defines the signalness as 

\begin{equation}
    S = \frac{N_\mathrm{signal} (E, \delta) }{ N_\mathrm{signal} (E, \delta) + N_\mathrm{background} (E, \delta)},
\end{equation}

with $N_\mathrm{signal} (E, \delta)$ and $N_\mathrm{background} (E, \delta)$ as the expected number of signal and background events with energy $E$ from declination $\delta$ based on simulations. From this, we get $N_\mathrm{signal} (E, \delta) / N_\mathrm{background} (E, \delta)) = S / (1-S)$ as our energy weights. The final weight is then the product of the spatial weights and the energy weights. Even though we do not limit the contributing events to the multiplets only, we expect the strongest contributions from the multiplet events because of their spatial proximity. 

Including these weights (following \citealt{2024JCAP...07..057K}), the likelihood describing the probability of observing an event at time $t_i$ given $K$ Gaussian flares ($\mathcal{N}$) and a uniform background becomes:

\begin{equation}\label{eq:em_likelihood}
    \mathcal{L} = \prod_i^N \left( \sum_k^K \frac{n_{\mathrm{S}, k}}{N} \mathcal{N}(t_i| \mu_{\mathrm{T}, k}, \sigma_{\mathrm{T}, k}) \frac{S_i}{B_i} + \frac{N - n_{\mathrm{S}}}{N} \frac{1}{(t_\mathrm{max} - t_\mathrm{min})} \right). 
\end{equation}

As before, $N$ is the total number of events, and $K$ is the maximal number of Gaussian contributions. Each Gaussian, $k$, is scaled by the associated number of signal events, $n_{\mathrm{S}, k} \geq 0$,  and each event probability is multiplied with the weight $S_i/B_i$. The rightmost term describes the uniform background component (becoming the sole component when setting $\sum_k^K n_{\mathrm{S},k} = n_\mathrm{S} = 0$). 

\begin{figure*}
    \centering
    \includegraphics[width=\linewidth]{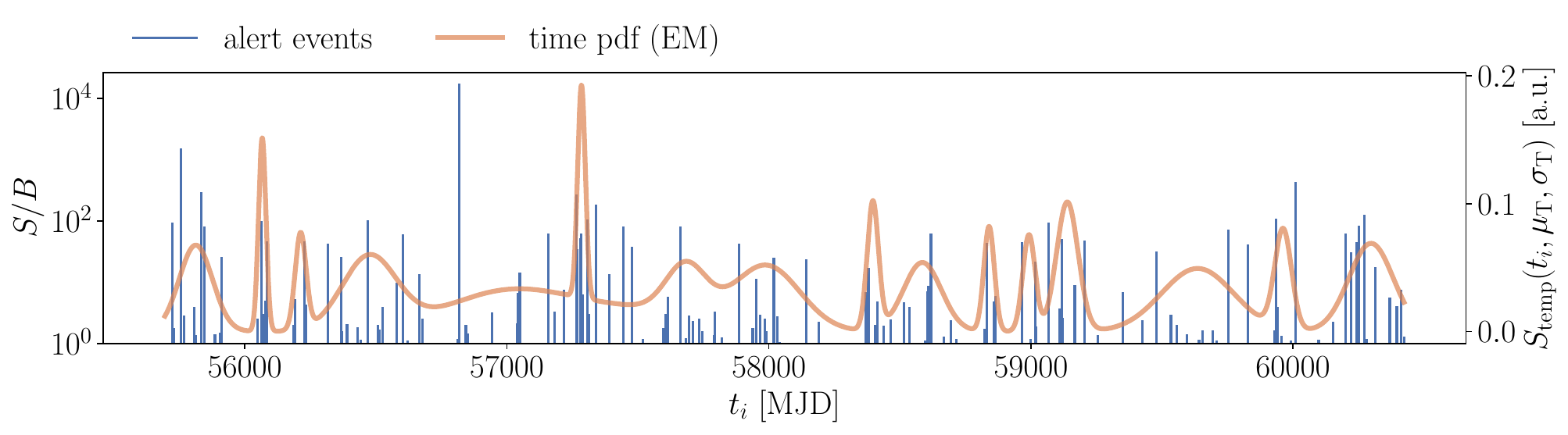}
    \caption{An example of a fitted time series. The left y-axis shows the applied weights for each event (Signal over Background: $S/B$). The x-axis shows the time range in MJD and the vertical blue lines indicate the detection times of the alert events. The orange line shows the best fit temporal signal probability density function, $S_\mathrm{temp}$, consisting of the sum of $k$ Gaussians, as determined with EM in arbitrary units. The height of each Gaussian is scaled with the number of associated events, $n_{\mathrm{S,} k}$. $S_\mathrm{temp}$ can reach greater values where many $n_{\mathrm{S,} k}$ contribute, compared to times where a single event has a high $S/B$ value without any close events in time. }
    \label{fig:time_series_example}
\end{figure*}

As described above, we assume a uniform background distribution in time and describe a neutrino signal by a set of normal distributions. We need to define some starting values as seeds for the optimization. First, we set the maximum number of flares to $K=20$. We choose this number to exceed the number of events in the biggest multiplet, which can serve as an indication of the upper bound of how many flares we can expect. Setting the number higher is unproblematic since some of these flares will be fitted to (close to) zero. We want to cover the full time period with the seed flares such that each event can contribute to the likelihood. Hence, we distribute the seed flares uniformly over the time series with a very broad width, $\sigma_\mathrm{T}$, of 500 days. For the beginning, we define an arbitrary flare strength ($n_{\mathrm{S,} k}$) of $\min(N / K - 1, 10)$ neutrinos, where we ensure that the sum of events belonging to flares does not exceed the total number of alert events: $\sum_k^K n_{\mathrm{S,} k} < N$. To avoid singularities (for $\sigma_\mathrm{T} \rightarrow 0$ in the $1/\sigma_\mathrm{T}$ term of the normal distribution), we require a minimal flare width of 10~days. We start running EM with these values and apply the same convergence criteria as in \citet{2024JCAP...07..057K, Abbasi_2024} (no change in the likelihood in the last 20 iterations or a maximum of 500 iterations). We show a fitted example time series of a multiplet in Figure \ref{fig:time_series_example}.

The resulting best-fit flares can then be used to calculate the likelihood (see equation \ref{eq:em_likelihood}) and further to calculate a test statistic value, by constructing a likelihood-ratio test. For the likelihood ratio test, the two hypotheses are as described above. The background  likelihood becomes the rightmost term of equation \ref{eq:em_likelihood} with $\sum_k ^K n_{\mathrm{S},k} = n_\mathrm{S} = 0$.

In order to determine if a temporal clustering is significant, we need to compare it with a background scenario. For this, we create random unclustered realizations of the sky by assigning randomized uniformly distributed right ascensions and detection times. We then repeat the described procedure on these random realizations. First, we identify the multiplets and calculate the weighted arithmetic circular mean positions. Then, we run EM with the same starting seeds as above for each position. This yields a number of multiplets, average positions, and best-fit Gaussian temporal flares for each background realization. However, the number and location of multiplets vary for each random realization, which makes a comparison of individual ``real'' multiplets to the background-generated ones not straightforward. 

Nonetheless, we can still compare the full sky of multiplets instead of looking at individual cases. We define a test statistic value for the full sky by summing up the results for all individual multiplets. This means, for every multiplet we get a result for the likelihood ratio test comparing the background and signal hypothesis. The test statistic for the full sky, $TS_\mathrm{temp}$,  is then the sum of the individual results ($TS_m$) over all multiplets $m$: $TS_\mathrm{temp} = \sum_m TS_m$. We repeat this for each sky realization and create a background $TS_\mathrm{temp}$ distribution. This follows the approach of stacking in \citet{Abbasi_2024}.

We find no significant temporal clustering of alert events independent of applied area, energy, or signalness thresholds. The best p-values for the area, energy, and signal searches are 0.35 (for $A \leq 6.27$~deg$^2$), 0.72 (for $E \geq 54$~TeV), and 0.93 (for $S \geq 0.414$ and $S \geq 0.452$). This allows several interpretations. It is possible that considering only the sum of all multiplets' test statistic values washes out a potential signal present in only few multiplets. Another possibility is that alert events are rare events in the time dimension and IceCube only detects a single event over 13 years from a source. This would require either a more sensitive detector or longer integration time in order to detect a clustering or multiple flares. 

\begin{figure}
    \centering
    \includegraphics[width=\linewidth]{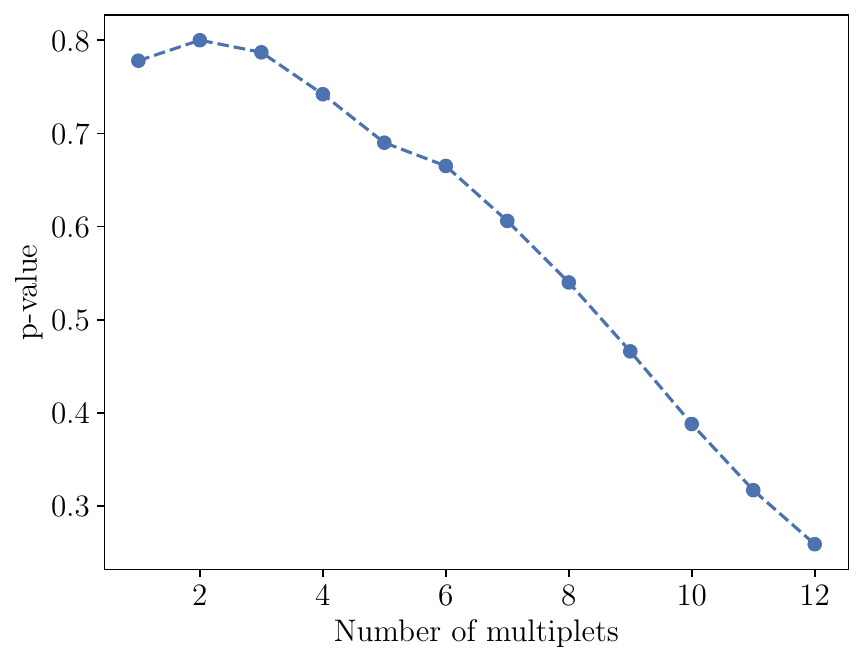}
    \caption{P-values when considering the top-ranked number of multiplets for an area threshold of $A \leq 6.27$~deg$^2$. The x-axis shows how many multiplets contribute to the p-value (sorted by highest test statistic value), and the y-axis shows the p-value. In this case, adding multiplets (by summing the test statistic values) lowers the chance probability, and we get the best p-value of 26\% when considering all multiplets. }
    \label{fig:binom_test}
\end{figure}

To check the first possibility (a signal of a few sources is washed out by looking at the sum of all multiplets), we test the most significant $m$ multiplets. We start with the multiplet yielding the largest test statistic value and compute the p-value by constructing a background test statistic distribution with the respective largest $TS$ values from the background trials. Then we take the sum of the two largest TS values and calculate their significance. We add more and more multiplets until we sum over every multiplet in the data. The last scenario differs slightly from the test in the previous paragraph (where we stack every multiplet in the sky) because here we always sum up to the maximum number of multiplets we found in the real sky. Even if we find more multiplets in the background sky, we do not consider more than the number of multiplets identified in the real sky. When running this test, we compare all p-values and get the most significant one (p=0.26) when applying an area threshold of $A \leq 6.27$~deg$^2$ and summing all (=~12) multiplets (see Fig. \ref{fig:binom_test}).  In this specific case, adding more multiplets lowers the p-value and increases the significance. This indicates that taking the sum of multiplets is not washing out a potential signal in this case. Interestingly, this behavior is not always repeated. When cutting on signalness, for example, the p-values rise for each threshold when including more multiplets. However, all p-values are still compatible with background. 

\section{Conclusions}

A fraction of alert events published by the IceCube Neutrino Observatory is expected to point to their cosmic production sites. We searched for an over-fluctuation of alert events with overlapping uncertainty regions (multiplets) compared to randomly distributed alert events. To reduce chance coincidences, we tested refined samples by restricting the uncertainty areas, the reconstructed neutrino energies, and their signalness. In all cases, we found no significant over-fluctuation of multiplets. We furthermore tested alternative reconstruction with smaller uncertainties (as published in the GCN Notices) but also there we find no significant clustering of events. A possible explanation could be that there are too many atmospheric background events included in the sample and our selections are not sufficiently signal-pure. Another possibility is that IceCube would either need a larger effective area or longer integration times to detect multiple alert events from the same source.

We then compared the number of $\gamma$-ray-detected blazars within the weighted mean area of each multiplet and found that they agree with the expected number of sources based on their average distribution. Hence, we do not find an excess of $\gamma$-ray detected blazars at the weighted mean area of the multiplets.

As a final step, we investigated the time series of each multiplet for the different thresholds in area, energy, and signalness. We allow for multiple flares fitted with expectation maximization and compare the resulting likelihood with randomly generated neutrino skies (random right ascensions and times). We then consider the sum of the resulting test statistic value over all tested positions. Combining all identified multiplets, we do not find a significant clustering in time. Possible reasons for this are that taking the sum over all multiplets washes out a signal only present in few locations or that alert events are detected too rarely to show a temporal clustering. We test the first case by starting with the most significant multiplet and then adding one by one less significant multiplets while observing how this affects the p-value. Doing this, we find a best p-value of 26\% for an area threshold of $A \leq 6.27$~deg$^2$ and considering all (in this case twelve) multiplets. This implies that, in this case, we do not wash out a signal when adding more multiplets. However, the behavior of improved p-values with more multiplets is not always repeated, which also shows that the result strongly depends on our selection of events. 

In any case, improving the number of alert events will allow a better understanding of whether those events cluster around their sources and could lead to the identification of neutrino sources. Furthermore, applying updated reconstruction contours \citep{2024GCN.38267....1I} on all previously published alert events will affect the uncertainty regions and will hence also influence a spatial correlation search. 

\begin{acknowledgements}
We thank Cristina Lagunas Gualda for the helpful discussion concerning the IceCube alert events and their reconstruction. We furthermore thank Chiara Bellenghi, Elena Manao, Rouhan Li, and Xavier Rodrigues for proofreading the manuscript. 
This work is supported by the Deutsche Forschungsgemeinschaft (DFG, German Research Foundation) through grant SFB 1258 ``Neutrinos and Dark Matter in Astro- and Particle Physics''.
\end{acknowledgements}

\bibliographystyle{aa} 
\bibliography{bibliography.bib} 

\end{document}